\documentclass[aps,pre, twocolumn,superscriptaddress,nopacs,amsmath,amssymb,final,letterpaper]{revtex4}

\usepackage[utf8]{inputenc}
\usepackage{calc}
\usepackage{graphicx}
\usepackage{amsmath,amssymb,amsthm}
\usepackage{txfonts}
\usepackage{bm}
\usepackage{color}
\usepackage{hyperref}
\usepackage{algcompatible}

\begin{document}
\author{Laurent H\'{e}bert-Dufresne}
\affiliation{D\'epartement de Physique, de G\'enie Physique, et d'Optique, Universit\'e Laval, Qu\'ebec (Qu{\'e}bec), Canada G1V 0A6}
\affiliation{Santa Fe Institute, Santa Fe, NM 87501, USA}
\author{Antoine Allard}
\affiliation{D\'epartement de Physique, de G\'enie Physique, et d'Optique, Universit\'e Laval, Qu\'ebec (Qu{\'e}bec), Canada G1V 0A6}
\affiliation{Departament de F\'isica Fonamental, Universitat de Barcelona, Mart\'i i Franqu\`es 1, 08028 Barcelona, Spain}
\author{Jean-Gabriel Young}
\author{Louis J. Dub\'{e}}
\affiliation{D\'epartement de Physique, de G\'enie Physique, et d'Optique, Universit\'e Laval, Qu\'ebec (Qu{\'e}bec), Canada G1V 0A6}

\title{On the constrained growth of complex scale-independent systems}

\begin{abstract}
Scale independence is a ubiquitous feature of complex systems which implies a highly skewed distribution of resources with no characteristic scale. Research has long focused on \emph{why} systems as varied as protein networks, evolution and stock actions all feature scale independence. Assuming that they simply do, we focus here on describing \emph{how} this behavior emerges, in contrast to more idealized models usually considered. We arrive at the conjecture that a minimal model to explain the growth towards scale independence involves only two coupled dynamical features: the first is the well-known preferential attachment principle and the second is a general form of delayed temporal scaling. While the first is sufficient, the second is present in all studied data and appears to maximize the speed of convergence to true scale independence. The delay in this temporal scaling acts as a coupling between population growth and individual activity. Together, these two dynamical properties appear to pave a precise evolution path, such that even an instantaneous snapshot of a distribution is enough to reconstruct the past of the system and predict its future. We validate our approach and confirm its usefulness on diverse spheres of human activities ranging from scientific and artistic productivity, to sexual relations and online traffic.
\end{abstract}

\maketitle

\section{Introduction}

Human systems are often characterized by \emph{extreme inequalities}. One may think of the distribution of wealth between individuals, the sizes of cities, or the frequencies of sexual activities to name a few \cite{Sornette, Champernowne, Zipf, Newman2005, Bettencourt07}. Interestingly, inequality often tends to manifest itself through a \emph{scale independent} behavior \cite{Sornette, Zipf, Champernowne, Newman2005, Bettencourt07, Yule2, Simon, Price, Barabasi1999, Hebert2011_prl, BTW, Bak, Carlson2000}. In layman's terms, these systems are said to be scale independent because of the absence of a characteristic scale. Taking the distribution of wealth as an example, the worldwide average income is meaningless because the variance is too wide. Neither the very poor nor the very wealthy can be reduced to average individuals; the former are too numerous while the latter are absurdly richer than the average.

\begin{figure*}
\centering
\includegraphics[width=\linewidth]{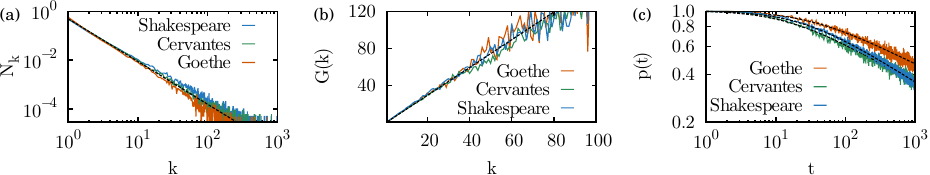}
\caption{(color online) (a) Power-law distribution of word occurrences in the writings of authors in three different languages. A power law with scale factor $\gamma = 1.75$ is plotted to guide the eye. 
Numerical scale exponents are estimated to be 1.89 for Goethe, 1.76 for Cervantes, and 1.67 for Shakespeare by the method of \cite{Clauset2007}. (b) Preferential attachment in written text with a linear relation for comparison. The algorithm to obtain 
the actual $G(k)$ is given in Appendix \S 4. (c) Average birth function for samples of 1000 words, this procedure is based on the translational invariance \cite{Bernhardsson2010} of written texts and yields better statistics. 
Three instances of Eq.~(\ref{pt}) are displayed with $\left[\alpha,\tau,b\right]$ equal to $\left[0.22, 31, 0\right]$, $\left[0.25, 15, 0\right]$ and $\left[0.28, 25, 0\right]$ [with $a$ fixed by $p(1)=1$] for Goethe's, Cervantes' and Shakespeare's writings respectively. This asymptotic scaling is related to what is generally known as Heaps' law of vocabulary growth in linguistics \cite{heaps}, but is given here a much more general expression for all $t$.}
\label{ProseResults}
\end{figure*}

Mathematically, this behavior takes the form of a \emph{power-law distribution}. That is, the number $N_k$ of individuals having a share $k$ (e.g. personal income or sexual partners) of the total resource $K$ (total wealth or sexual activities) roughly follows $N_k \propto k^{-\gamma}$. One of the first robust observation of scale independent systems concerns the distribution of occurrences of individual words in prose \cite{Zipf} as illustrated in Fig. \ref{ProseResults}(a). %In this relation, $\gamma$ is said to be the \emph{scale parameter} because any change $\lambda$ in the scale of interest can be written as $n(\lambda k) \propto \lambda ^{-\gamma} n(k)$.

In this paper, we build upon two general premises to describe the growth of scale independent systems. Firstly, we assume that the underlying distribution roughly follows $N_k \propto k^{-\gamma}$ such that a power law is an adequate approximation for sufficiently large $k$ (with $\gamma > 1$ for normalization in the asymptotic limit). Secondly, we follow the distribution of a resource or property that can only increase or stagnate, namely the total activities of an individual (both past and present).

Throughout the paper, time and system size (in terms of the resource $K$) are completely interchangeable. This stems from the fact that our description of a complex system is usually based on a fixed dataset with no temporal information. By considering the dataset as an underlying growing system, to which we do not have access, the only available notion of time is the number of entries. These entries assign a new unit of the resource $(K(t)=K(t-1)+1)$ to one of the $N(t)$ individuals. Based on our simple assumptions, the resulting model will be able to constrain the probabilities of various future entries in the actual dataset.

The paper is organized as follows. In Sec.~\ref{sec:model}, we construct our theoretical framework and obtain a versatile minimal growth model, a generalization of the standard preferential attachment approach. In Sec.~\ref{sec:results}, we use diverse databases to validate our method: scientific productivity of authors on the arXiv e-print archive (arXiv), one month of user activities on the Digg social news website (Digg) \cite{Lerman10}, productivity of actors on the Internet Movie Database (IMDb) and sexual relations in a Brazilian escort community (sexual) \cite{Holme2011}. Based on the successes of these empirical evidences, we thereby confirm that our framework can be used not only to infer the past of known distributions but to construct their future. We conclude, in Sec.~\ref{sec:conc}, by summarizing what insights and applications are offered by our work. Some technical details of the analysis and a description of our datasets and algorithms are relegated to a separate Appendix.

\section{Theoretical Framework\label{sec:model}}

Let us consider the growth of a hypothetical system where each individual $i$ possesses a share $k_i(t)$ of the total resource $K(t)$ at time $t$. Because the system is constantly growing, both in terms of its total population $N(t)$ and of each individual's share, time can be measured as the total number of events. These events can take one of two forms: \emph{birth events} which increase the total population $N(t+1) = N(t)+1$ by adding a new individual $j$ with $k_j(t) = 1$; and \emph{growth events} which imply $k_i(t+1) = k_i(t)+1$ for a given individual $i$.

We then introduce two functions: a \emph{birth function} $p(t)$ that prescribes the probability that the $t$-th event is a birth event, and a \emph{growth function} $G(k)$ that describes the average chances (unnormalized probability) for an individual with current share $k$ of being involved in the next growth event. Assuming that individuals with the same share are indiscernible, the state of an average individual $i$ of share $k_i$ can be followed through a mean-field model:
\begin{equation}
k_{i}(t+1) = k_i(t) + \left[1-p(t)\right]\frac{G\left(k_i(t)\right)}{\sum _j G\left(k_j(t)\right)} \;.
\label{k}
\end{equation}
Consequently, the probability that a growth event involves \emph{any} individual of current share $k$ is given by $N_k(t)G(k) / \sum _{k'} N_{k'}(t)G(k')$ where $N_k(t)$ is the number of individuals with share $k$ at time $t$. This yields the following master equation (for $k$ $\in$ $\mathbb{N}$)
\begin{align}
N_k(t+1)  =  &N_k(t) + p(t)\delta_{k,1}\nonumber \\ &+ \left[1 - p(t)\right]\dfrac{N_{k - 1}(t)G(k - 1) - N_k(t)G(k)}{\sum _m N_m(t)G(m)}  \label{model}
\end{align}
with $N_0(t)=0$ $\forall t$. For this model to be of any use, at least partial knowledge of $G(k)$ and $p(t)$ is required. Setting $G(k)=k$ and a constant $p(t)$, we retrieve the \emph{classic preferential attachment} process \cite{Simon}. However, our goal is to investigate the constraints imposed by the scale independence, $N_k(t) \propto k^{-\gamma}$, on the functional forms of both $p(t)$ and $G(k)$ as well as the coupling between the two.

The next two sub-sections are more technical in scope, but necessary to delineate the functional forms that will constitute the basis of the studies presented in section \ref{sec:results}. Although our analysis is based on asymptotic arguments, and therefore approximate, we will demonstrate that the following expression, 
\begin{equation}
p(t) = a(t+\tau)^{-\alpha} + b \;
\end{equation}
combining three adjustable parameters $[\alpha, \tau, b]$ ($a$ can be removed by normalization), together with $G(k) \to k$ and the 
dynamical model of Eq. (\ref{model}), captures the essence of the growth of diverse human activities. The form of $G(k) \propto k$, at least for $k$ greater than a certain bound $k^{*}$, is not new, but emerges naturally from our premises. As we will see shortly, the temporal dependence of $p(t)$ is inherent to the growth towards scale independence and is coupled to the behavior of $G(k)$ at small $k<k^{*}$ through the parameter $\tau$.

\subsection{The growth function}
%We define the growth function $G(k)$ as the chances (unnormalized probability) for an individual with share $k_i(t) = k$ to see its share increase during the next time step. 

The behavior of the growth function $G(k)$ can be constrained by an argument presented by Eriksen and H\"ornquist \cite{eriksen01}. We wish to obtain $G(k)$ solely on the basis of Eq.~(\ref{model}). Instead of measuring $G(k)$ directly by looking at what leaves the compartment $N_k(t)$, we can equivalently look at what arrives in the compartments $k'>k$ during the time step $t \rightarrow t+1$. We write this as the difference between what is in $k'>k$ at $t+1$ [i.e. $\sum _{i=k+1}^{\infty} N_i(t+1)$] and what was in $k'>k$ at time $t$ [i.e. $\sum _{i=k+1}^{\infty} N_i(t)$]. We substitute $N_i(t+1)$ with Eq.~(\ref{model}) and sum over all $k'>k$:
\begin{align}
\sum _{i=k+1}^{\infty} & \left[N_i(t+1)-N_i(t)\right] \nonumber \\ & = \sum _{i=k+1}^{\infty} \left\{ p(t)\delta_{i,1} + \left[1 - p(t)\right]\dfrac{N_{i - 1}(t)G(i - 1) - N_i(t)G(i)}{\sum _m N_m(t)G(m)} \right\} \nonumber \\
 & = \left[1-p(t)\right]\frac{N_k(t)G(k)}{\sum _m N_m(t)G(m)} \; .
 \label{dummy}
\end{align}
This last expression can be interpreted as two measures of the activity in compartment $N_k(t)$ between $t$ and $t+1$. The left-hand side measures the mean number of arrivals in compartment $N_{k'}(t)$ with $k'>k$; i.e. the mean number of individuals which left compartment $N_k(t)$. The right-hand side is explicitly the ratio of the activity involving the $k$-th compartment, $N_k(t)G(k)$, to the total growth activity, $\sum _m N_m(t)G(m)$, times the probability, $1-p(t)$, that a growth event has occurred during the time step. From this equivalence, $G(k)$ is readily obtained from Eq. (\ref{dummy}):
\begin{equation}
G(k) = \frac{\sum _m N_m(t)G(m)}{1-p(t)}\frac{1}{N_k(t)} \sum _{i=k+1}^\infty \left[N_i(t+1) - N_i(t)\right] \; .
\label{Gk}
\end{equation}
For $k\gg 1$, we can replace the sum by an integral, and using our only hypothesis, i.e. $N_k(t) = A(t) k^{-\gamma} N(t)$, where $A(t)$ is a normalization factor, we find
\begin{align}
G(k) \simeq & \frac{\sum _m N_m(t)G(m)}{1-p(t)}\left[\frac{A(t+1)N(t+1) - A(t)N(t)}{A(t)N(t)}\right] \frac{k}{\gamma - 1} \; .
\end{align}
All factors independent of $k$ are of no concern, since $G(k)$ only makes sense when comparing the relative values for different $k$. Hence, at any given time $t$, we finally obtain
\begin{equation}
G(k) \propto k
\end{equation}
at least for values of $k$ higher than an appropriate lower bound. This linear relation between the probability of growth of an individual and its present size, \emph{preferential attachment}, is a recurrent feature in scale independent growth models \cite{Yule2, Gibrat, Champernowne, Simon, Price, Barabasi1999, Hebert2011_prl}. This simple derivation states once again that a scale independent growing system implies a linear preferential attachment. See Fig. \ref{ProseResults}(b) for examples. However, observing preferential attachment in datasets do not imply that preferential attachment is the active mechanism in the growth process, but simply that past activity is at least correlated with whatever growth mechanism is actually at play. One should then think of preferential attachment as an \textit{effective mechanism} that reproduces the \textit{statistical} properties of growth. This statement becomes particularly relevant when one considers for instance the writings of William Shakespeare, Miguel de Cervantes Saavedra and Johann Wolfgang von Goethe analysed in Fig. 1. No one in their right mind would consider
preferential attachment as the operational mechanism governing the authors' choices of words, even if its statistical signature is present.

In recent years, the idealized preferential attachment process, using $G(k) = k$ and $p(t) = p$, has been analysed to great lengths. Most studies have been concerned with the application of this process to network growth \cite{Doro_evolution, Albert_RMP} and have focused on solving the resulting network structure \cite{Redner_prl00, Doro_prl00}, describing the statistics of leading nodes \cite{Redner_prl02}, finite-size effects \cite{Bagrow08}, and its relation to other properties of complex networks such as their modular and self-similar nature \cite{lhd_pre12}.

\subsection{The birth function\label{Sec_1B}}

A time-varying birth rate $p(t)$ has been considered before, either in \emph{ad hoc} manner \cite{Simon, Zanette} or in a specific context \cite{Gerlach13} based on empirical observations in, for example, written texts \cite{heaps} or human mobility \cite{Song10}. Instead of investigating how a given $p(t)$ might influence the distribution of resource in the system, we investigate how a given distribution of resource informs us on the actual $p(t)$ of that system. In doing so, the hope is to provide a more general framework for understanding how and why scale independent organization implies scale independent growth.

In our model, the birth function has two important roles. First, it is equivalent to the time derivative $\dot{N}(t)$ of the population $N(t)$; and second, it constrains the growth of the largest share $k_{\textrm{max}}(t)$. Two relations can be called upon to connect $N(t)$ and $k_{\textrm{max}}$, and obtain a consistent functional form for $p(t)$.%There exist two relations to connect $N(t)$ and $k_{\textrm{max}}(t)$, such that a functional form of $p(t)$ can be obtained. 

The first relation is the extremal criterion \cite{Redner_prl00}: $\int _{k_{\textrm{max}}(t)}^\infty N_{k}(t)dk \sim 1$, intuitively meaning that the number of individuals with a maximal share is of order one. To simplify the analysis, we will assume that $k_{\textrm{max}}(t) \gg 1$, such that the normalization $A(t) = \left[\sum _1^{k_{\textrm{max}}(t)} k^{-\gamma}\right]^{-1}$ has converged to a constant $A^*$. We thus use $N_k(t) = A^ *N(t)k^{-\gamma}$ in the extremal criterion and solve for $N(t)$:
\begin{equation}
N(t) \sim \frac{\gamma - 1}{A^*}k_{\textrm{max}}^{\gamma-1}(t) \quad \rightarrow \quad \frac{N(t)}{\dot{N}(t)} = \frac{k_{\textrm{max}}(t)}{\left(\gamma-1\right)\dot{k}_{\textrm{max}}(t)} \; .
\label{k_from_N}
\end{equation}
Note that keeping the temporal dependency of $A(t)$ yields the same result for the leading temporal term. The second important relation stems from our definition of time $t$ (in number of events or resource $K$) such that $\dot{K}(t) = 1$. We write
\begin{align}
\dot{K}(t) =& \frac{d}{dt}\sum _{m=1}^{k_{\textrm{max}}(t)} m N_m(t) \nonumber \\ =& \frac{d}{dt}\left[\sum _{m=1}^{k^*} m N_m(t) + \int _{k^*}^{k_{\textrm{max}}(t)} mN_m(t)dm\right] = 1
\end{align}
where $k^*$ is an appropriate bound for the integral approximation of the sum. Again, using $N_k(t) = A^ *N(t)k^{-\gamma}$, we obtain
\begin{equation}
A^*\dot{N}(t)\left[C + \frac{1}{2-\gamma}k_{\textrm{max}}^{2-\gamma}(t) + \frac{N(t)}{\dot{N}(t)} k_{\textrm{max}}^{1-\gamma}(t)\dot{k}_{\textrm{max}}(t)\right] = 1 \; ,
\end{equation}
where $C$ is a constant collecting all terms independent of $t$. Replacing $N(t)/\dot{N}(t)$ with Eq.~(\ref{k_from_N}) allows us to solve for $\dot{N}(t)$ [i.e. $p(t)$]:
\begin{equation}
p(t) = \dot{N}(t) =\frac{\left(2-\gamma\right)\left(\gamma-1\right)}{A^*\left[C\left(2-\gamma\right)\left(\gamma-1\right)+k_{\textrm{max}}^{2-\gamma}(t)\right]}
\end{equation}
If $\gamma \in (1,2)$, $k_{\textrm{max}}^{2-\gamma}(t)$ is the leading term and $p(t)$ decreases as $k_{\textrm{max}}^{\gamma-2}(t)$; if $\gamma > 2$, $k_{\textrm{max}}^{2-\gamma}(t)$ becomes negligible and $p(t)$ is essentially governed by the first two terms of the ensuing geometric series. We can summarize these results, obtained only by assuming $N_k(t) \propto k^{-\gamma}$ and $k_{\textrm{max}}(t) \gg 1$, under a general form
\begin{align}
p(t) \propto &
\begin{cases}
k_{\textrm{max}}^{\gamma-2}(t) & \textrm{if } \ 1 < \gamma < 2 \\
k_{\textrm{max}}^{2-\gamma}(t) + \textrm{constant} & \textrm{if } \qquad \gamma > 2 \; .
\end{cases}
 \label{pt0a}
\end{align}
%or using the following functional form:
%\begin{equation}
%p(t) = a_0k_{\textrm{max}}(t)^{-\alpha_0} + b_0\; 
%\label{pt0}
%\end{equation}
%with appropriate free parameters. 

\begin{figure*}
\centering
\includegraphics[width=0.75\linewidth]{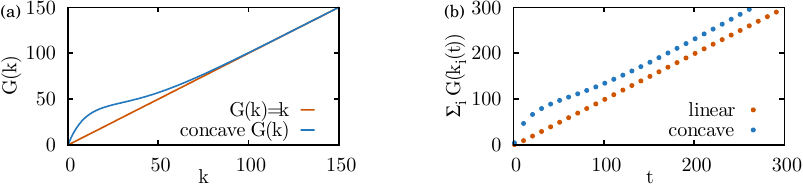}
\caption{(color online) Two different growth functions (a), the classic $G(k) = k$ and a concave $G(k) = k + k\,\textrm{exp}(-k/15)$, and their effect on the total growth chances $\sum _i G(k_i(t))$ (b). The non-linearity in the second growth functions is reproduced in the time evolution of the total system. During the early stages of the dynamics, most events occur at small $k$ where $G(k+1)-G(k) > 1$, causing the sum to grow faster than expected from the asymptotic linearity of $G(k)$. Consequently, even though $\sum _i G(k_i(t))$ converges to a linear behavior for large $t$, an offset (i.e., $\kappa\tau$) remains to account for the initial non-linearity. The temporal results (b) are obtained by iterating Eq. (\ref{model}) with $p(t) = 0.01$, and the observed offset $\kappa\tau$ is in perfect agreement with the results of the Appendix \S 1.}
\label{SumResults}
\end{figure*}

\begin{table}[h!]
\caption{Definitions of important functions and parameters.}
%\vspace{-0.5cm}
\begin{center}
\begin{tabular}{cl} \hline\hline 
$N_k(t)$ & number of individuals of share $k$ of a total resource $K$\\
             & {\em assumed} scale independent $\propto k^ {-\gamma}$ for large $t$\\
              \hline \hline
$G(k)$ & {\em growth function}: chances that a growth event \\
             & involves an element $i$ with share $k_i(t)= k$ \\ 
             & \\
             $\kappa$& multiplicative factor of the delayed linear scaling \\
             & of the normalization of $G(k)$, i.e., $\sum_i G (k_i(t)) \simeq \kappa(t+\tau)$ \\ 
             \hline\hline
$p(t)$ & {\em birth function}: probability that the $t$-th event  \\
             & is a birth event \quad $p(t) = a (t+\tau)^{-\alpha} +b$ \\ 
             & \\
             $\alpha$ &   temporal scaling \\
             $\tau$ & temporal delay caused by non-linearity in $G(k)$ \\
              $b$ & asymptotic value \\
              $a$ & normalization \\
              
              \hline \hline             
\end{tabular}
\end{center}
\label{table0}
\end{table}

The remaining step is to establish the time dependence of $k_{\textrm{max}}(t)$ to obtain the explicit temporal form of $p(t)$. In line with our asymptotic arguments, as $k_{\textrm{max}}(t)$ increases beyond an appropriate bound $k^*$ 
where $G(k) = k$,  Eq. (\ref{k}) simplifies to
\begin{equation}
k_{\textrm{max}}(t+1) = \left[1+\frac{1-p(t)}{\kappa\left(t+\tau\right)}\right]k_{\textrm{max}}(t) \; .
\label{kmax}
\end{equation}
The denominator represents the asymptotic behavior of the normalization of growth probabilities
$\sum_k G(k) N_k(t)$ which can be shown to converge to $\left[\kappa (t+\tau)\right]$ for $t\gg 1$.  The derivation of this result and the expressions for the constant $\kappa$ and the delay $\tau$ are
presented in Appendix \S 1. The initial and arbitrary behavior of $G(k)$ offsets the value of the sum by a constant expressed
as a temporal delay $\kappa\tau$.

This offset is illustrated in Fig.~\ref{SumResults} for two different growth functions.
%While we have shown that $G(k) = c_1 k$ for $k>k^*$, the parameter $\tau$ is introduced to account for the potential non-linear behavior of $G(k)$ for $k < k^*$. More precisely, it can be verified (analytically or numerically) that the sum normalizing the growth probabilities, i.e. $\sum _k G(k)N_k(t)$, converges to a linear function of time irrespectively of the initial behavior of $G(k)$. This behavior can however offset the value of the sum by a constant, $c_1 \tau$, which is positive (negative) when $G(k)$ is greater (smaller) than $c_1 k$ for $k<k^*$. 

Equation (\ref{kmax}) determines the derivative in the limit of large $t$,
\begin{equation}
\frac{d}{dt}k_{\textrm{max}}(t) = \frac{1-p(t)}{\kappa\left(t+\tau\right)}k_{\textrm{max}}(t) \; .
\label{eq:diff_kmax}
\end{equation}

Since $p(t)$ is limited to the range $[0,1]$ we can write, without loss of generality, $p(t) = f(t) + b$ where $b$ is the asymptotic value of $p(t)$. This form yields the exact solution
\begin{equation}
k_{\textrm{max}}(t) = C_1 (t+\tau)^{(1-b)/\kappa}\textrm{exp}\left[-\int _{t^*}^{t} \frac{f(t')}{\kappa\left(t'+\tau\right)}dt'\right] \; 
\end{equation} 
where $t^*$ is an appropriate lower bound such that Eq.~(\ref{eq:diff_kmax}) is applicable. Since $f(t)$ is bounded, the exponential factor converges rapidly to one and we find the general solution for large $t$
\begin{equation}
k_{\textrm{max}}(t) = C_1 (t+\tau)^{(1-b)/\kappa} \; .
\label{kmaxf}
\end{equation}
Inserting Eq. (\ref{kmaxf}) in Eq. (\ref{pt0a}), we obtain a functional form for the birth function (with parameters summarized in Table. \ref{table0})
\begin{equation}
p(t) \simeq a\left(t+\tau\right)^{-\alpha} + b \; ,\label{pt}
\end{equation}
where we identify 
\begin{align}
\alpha = &
\begin{cases}
(2-\gamma)/\kappa & \textrm{if } \ 1 < \gamma < 2 \\
\left(\gamma-2\right)\left(1-b\right)/\kappa & \textrm{if } \qquad \gamma > 2 \, .
\end{cases}
 \label{alpha-gamma}
\end{align}

The first confrontation of Eq.~(\ref{pt}) with empirical data is displayed in Fig. \ref{ProseResults}(c). %Integrating Eq.~(\ref{pt}) to determine $N(t)$ and obtain $k_{\textrm{max}}(t)$ from Eq.~(\ref{k_from_N}) confirms the self-consistency of our results as we fall back on Eq.~(\ref{kmaxf}).

Before we describe in the next section the procedure adopted to fit the parameters $[\alpha, \tau, b]$ (the parameter $a$ is fixed by population size) on actual data, a few comments appear necessary. These three free parameters \emph{do not} overparameterize the function. Two of them, $\alpha$ and $b$, govern the scale exponent in the two fundamentally different regimes $\gamma < 2$ and $\gamma > 2$ respectively, while the delay $\tau$ embodies an intrinsic coupling between population growth and individual growth. For instance, as our results will illustrate, a large value of $\tau$ expresses the fact that the system features strong diminishing returns on growth for small $k$ (concave $G(k)$). To a lesser extent, $\kappa$ plays a similar role, although it is also coupled to other temporal ($b$) and organizational ($\gamma$) features within $\alpha$. 

From the asymptotic nature of our derivation, it is not to be expected that the relations of Eq.~(\ref{alpha-gamma}) between the exponents $\alpha$ and $\gamma$ should be strictly observed. However, the results of Fig. \ref{ProseResults} %(see the numerical values in the caption) 
indicate that it is nearly true for the three prose samples studied.   
These turn out to be cases with $b = 0$ and $\kappa = 1$ according to Eq.~(\ref{eqn:kappa}).
The values of $\alpha = 0.22, 0.25, 0.28$ and the corresponding inferred values of $\gamma= 2-\alpha = 1.78, 1.75, 1.72$ 
are indeed close to the numerical estimates of the scaling exponents, $\gamma = 1.89(4), 1.76(3), 1.67(8)$ respectively, obtained independently with the method of \cite{Clauset2007} .

For the cases where $b \neq 0$, the classical preferential attachment (CPA) limit [$G(k) = k$ and $p(t) = b$] of our model dictates that the asymptotic scaling exponent should be $\gamma_{\textrm{CPA}} = (2 - b)/(1 - b)$. Since the data will seldom have reached their asymptotic regime, deviations will be recorded and the connection between $\alpha$ and $\gamma$ will be partly lost. Moreover, to obtain asymptotic results for growth functions that are not strictly linear for all values of $k$, one must study each scenario on a case-by-case basis \cite{Redner_prl00, Doro_prl00}; estimating $\kappa$ alone requires the integration of the model. Nevertheless, despite the absence of exact expressions for $p(t)$ and $G(k)$, the flexibility of the derived functional form will provide a useful and versatile parametrization of the complete temporal evolution of empirical data. The results of the next section confirm this assertion.

%Using $G(k) = k$, this complete model yields $\alpha = 2-\gamma$ with $b=0$ when $1<\gamma<2$, and recovers the result of the classic preferential attachment\cite{Simon} when $\gamma > 2$ and $p(t) = b$, leading to the asymptotic scaling exponent $\gamma _{\textrm{PA}}= \left(2-b\right)/\left(1-b\right)$. These relations provide starting points for the application of our model{\color{red}, but are not strictly respected as more general growth functions have to be studied on a case-by-case basis\cite{Redner_prl00, Doro_prl00}. That being said, we now consider our main result, Eq. (\ref{pt}), as a useful parametrization of scale-independent systems. While we do not know the exact analytical connection between all parameters for general $p(t)$ and $G(k)$, our next results show that their functional forms remain valid and useful.}

%{\color{red} Finally, as previously mentioned, the delay $\tau$ accounts for both potential initial conditions and/or non-linear behavior of $G(k)$ at small values of $k$. It embodies an intrinsic coupling between population growth and individual growth. More precisely, we find that $\tau$ is very large when the system features strong diminishing returns on growth for small $k$.}

\section{Results\label{sec:results}}

The model based on Eq.~(\ref{model}) may now be used to replicate the growth of empirical distributions. Our objective is in part to verify the presence of constraints on the birth, Eq.~(\ref{pt}), and growth, Eq.~(\ref{Gk}), of individuals; but also to use them to determine the past and future of different systems solely from a snapshot of their present distribution.

\subsection{Reconstructing the past}

Our model consists of iterating Eq.~(\ref{model}) for all $k$, with a given combination of $p(t)$ and $G(k)$, until time $t$ reaches the total resource, $K$, of the system's present state. Hereafter, \textit{we do not at any point show actual fits of the temporal data}, but instead find the optimal combination of $p(t)$ and $G(k)$ that minimizes the error produced by Eq.~(\ref{model}) when modeling the present state of a given system. %The error produced by the model is then quantified by the number of misplaced individuals; i.e. the sum of differences between the actual distribution and the distribution of $N_k$ for all $k$.

A simple analogy will clarify first the strategy behind our optimization procedure. We are given a semi-infinite vertical chain of buckets. At the bottom of each one we drill a small hole of various width such that the $k$-th bucket has a hole of size $G(k)$. The first bucket, at the top of the chain, is placed under a dripping faucet whose flow is controlled in time by the function $p(t)$. Our goal is to adjust both the flow of the water $p(t)$ and the width of the holes $G(k)$ in order to reach a target quantity $\tilde{N}_k(t_f)$ of water for each bucket $k$ after a time $t_f$. This target quantity is itself produced by a hidden $\tilde{p}(t)$ and $\tilde{G}(k)$. Since the function $G(k)$ has an infinite number of degrees of freedom, this means that for almost any $p(t)$ we could find a $G(k)$ respecting the target distribution. However, if the chosen $p(t)$ is very different from $\tilde{p}(t)$, the obtained $G(k)$ will also differ from $\tilde{G}(k)$. Therefore, we constrain $p(t)$ first, having a few degrees of freedom, before optimizing $G(k)$ accordingly.

%We initially set $G(k) = k$ and we evaluate the quality of our model for the evolution of the system as obtained with different choice of $p(t)$. As shown in Sec. \ref{Sec_1B}, initially assuming $G(k) = k$ does not limit the range of reproducible scale exponents $\gamma$; all scale independent distributions can be modeled with appropriate $p(t)$. In the worst case, $p(t)$ would have four degrees of freedom, but we can remove one of them with our knowledge of the final population (or total quantity of water) which sets the temporal average $\langle p(t) \rangle$ = $\sum _k \tilde{N}_k(t_f) /t_f$. We can further reduce the number of possible parameters through simple conditions; for instance $p(t) \in [0,1]$. For each tested $p(t)$, we use the previously defined model (see Eq. (\ref{model})). 

The quality of our model representation $[p(t), G(k)]$ is assessed by counting the number of individuals $\{N_k(t_f)\}$ (or water drops) assigned to the wrong share $k$ (or the wrong bucket) with respect to the empirical state $\{\tilde{N}_k(t_f)\}$, 
\begin{equation}
\Delta \left[p(t),G(k)\right] =  \frac{1}{2}\sum _k \vert \tilde{N}_k(t_f) - N_k(t_f)\vert \; .
\label{epsilon}
\end{equation} 
A number of points are worth mentioning. Firstly, the measure $\Delta$, based on absolute errors, was chosen over, say logarithmic or cumulative errors, because of its robustness to the tails of the distributions where the finite-size data falls to a non-zero value ($\propto N(t_f)^{-1}$) while the mean-field model falls to zero. Secondly, although minimisation of $\Delta$ (or optimisation of $[p(t), G(k)]$) is conducted on the sole knowledge of the present state of the system, i.e. $\{\tilde{N}_k(t_f)\}$, our model completely reconstructs its pre-history. Thirdly, while the search for the optimal parameter values of $p(t)$ seems a daunting enterprise, a number of internal and empirical restrictions on $p(t)$ constrains the problem: i. since $p(t) \in [0,1]$ $\forall$ $t$, $b \in [0,1]$ and therefore $-b \leq a(t+\tau)^{-\alpha} \leq (1-b)$; ii. since $p(t) = \dot{N}(t)$ by definition, the total empirical population $\tilde{N}(t_f)$ can serve as normalisation, removing one degree of freedom:
\begin{equation}
a = \frac{\tilde{N}(t_f) - bt_f}{\left(t_f+\tau\right)^{1-\alpha}-\left(1+\tau\right)^{1-\alpha}}\left(1-\alpha\right)\; .
\end{equation}
Because $a$ can be positive or negative, our model can just as well describe a growing or decreasing birth function. Finally, the optimisation procedure is carried out in two stages: i. an initial set of optimal triplets $[\alpha, \tau, b]$ is obtained by scanning parameter space to minimize $\Delta$ while maintaining initially $G(k) = k$; % (Fig. \ref{sexualerror} presents an example of this parameter scan); \\
ii. the growth function $G(k)$ is then allowed to vary under the newly acquired best possible $p(t)$ and constrained by the empirical data $\{ \tilde{N}_k(t_f)\}$. Details of the algorithm are given in Appendix \S 5. Based on the quality of the obtained model $[p(t), G(k)]$, no further optimization was found necessary.

\begin{figure*}
\begin{center}
\includegraphics[width=\linewidth]{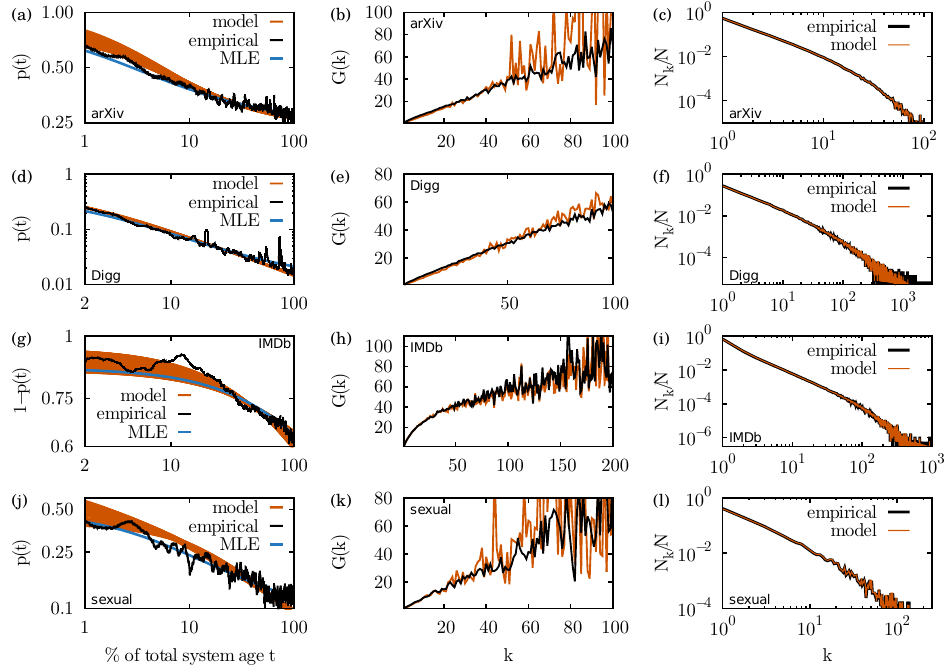}
\caption{(color online) \label{PastResults} From left to right: birth function with temporal scaling of the form $a(t+\tau)^{-\alpha}+b$; growth function with asymptotic preferential attachment; scale independent distributions. (a, d, g and j) The orange curves represent birth functions leading to predictions within $25\%$ of the minimal error between model and empirical data using the present state only. The empirical black curves are presented solely for comparison as no temporal data is needed for our reconstruction of the past. Likewise, Maximal-Likelihood Estimates (MLE) of $p(t)$, calculated \textit{with} the actual sequence of birth and death events are shown in blue to highlight the accuracy of our model. (b, e, h and k) Growth functions and (c, f, i and l) present distributions: only the curves with the absolute minimum error are shown. The systems are: (a, b and c) distribution of papers per author in the arXiv [$N(t_f) = 386,267$ at $t_f=1,206,570$], (d, e and f) votes per user on Digg [$N(t_f) = 139,409$ at $t_f=3,018,197$], (g, h and i) movies per actor on IMDb [$N(t_f) = 1,707,525$ at $t_f=6,288,201$] and (j, k and l) relations per individual in the sexual data [$N(t_f) = 16,730$ at $t_f=101,264$]. The methodology to measure the empirical birth and growth functions is presented in Appendix \S 3 and 4. }
\end{center}
\end{figure*}

While the systems studied in Fig.~\ref{PastResults} vary in nature, age and distributions, our results indicate that they follow qualitatively the same evolution, and confirm the presence of both a delayed regime of temporal scaling and preferential attachment in all cases.  Point estimates (Maximum-Likelihood Estimation (MLE) over the binary sequence of birth and growth events, see Appendix 
6) of the relevant parameters are given in Table \ref{table1} and are visually compared with our model in Fig.~\ref{PastResults}(a, d, g and j). The behaviors extracted by our model from static distributions (without temporal data) are thus shown to be good estimates of the best possible fits to the actual temporal data.
\begin{table}[h!]
\begin{center}
\caption{MLE point estimates of parameters using the empirical sequence of birth and growth events.}
\begin{tabular}{ @{}c c c c c@{} } \hline \hline
system & arXiv & Digg & IMDb & sexual \\
\hline
\hline
$\alpha$ & $0.58$ & $0.95$ & $0.46$ & $0.60$\\
$\tau$ & $12,066$ & $60,364$ & $6,288,202$ & $3,038$ \\
$b$ & $0.240$ & $0.012$ & $0.976$ & $0.072$\\
\hline
\hline
\label{table1}
\end{tabular}
\end{center}
\end{table}

%In the IMDb, death events (removal of certain individuals) could cause $\dot{N}(t)$ to be negative. Interestingly however, this situation appears to affect neither the functional form of $p(t)$, nor the efficiency of our model.

Because of the form $p(t) = a(t+\tau)^{-\alpha} + b$, the complementary probability (i.e. the probability that the $t$-th event is a growth event) has the same form with $a' = -a$ and $b' = 1-b$. This fact is highlighted with the case of IMDb in Fig. \ref{PastResults} and is consistent with our analysis where the constant $a$ (but not $b$) can be negative. Furthermore, notice that IMDb is not only the sole system for which $p(t)$ is an increasing function, but also the only system for which $G(k)$ has initially a non-linear behavior, and consequently a large $\tau$. This confirms our interpretation of the role of $\tau$ as a coupling between population growth, $p(t)$, and individual growth, $G(k)$. With hindsight, this initial regime of the IMDb growth function probably corresponds to the so-called \emph{star system}: actors with little experience are far less likely to be chosen for a role than experienced actors, but the first few movies in a new actor's curriculum are also far more important than the $n$-th in the career of a well-established star. This influences the introduction rate of new actors to preserve the system's scale independence. This interpretation is somewhat speculative, yet the fact remains that these effects are observed in the temporal data and that our model is able to extract them solely from the present distribution.

With the exception of one much smaller system (sexual data), the quality of our reconstruction of the past is surprisingly good considering that it requires no temporal data whatsoever. For instance, the Digg user activity distribution led us to determine with very high precision that 25\% of votes are due to new users 12 hours into the month, whereas this proportion falls below 2\% by the end of the month.

Our ability to infer the birth function based on a single snapshot also implies that we can distinguish between systems close or far from equilibrium (i.e. their statistical steady-state). For all investigated cases, both the inferred and observed $p(t)$ agree that none of these systems have reached their asymptotic $b$ value. In the Digg database, it is even unclear if this value exists at all. In other systems, it is interesting to discern whether the distribution is approaching its asymptotic scale exponent $\gamma$ from above (less heterogeneity) or below (more heterogeneity). For instance, the sexual database describes a network for which the first two moments of the activity distribution determine whether or not the introduction of a given sexually transmitted infection will result in an epidemic \cite{newman02, hebert10}. These moments being defined by the scale exponent, our ability to describe the system's approach to equilibrium directly translates in an ability to determine which infection could invade the network.

More generally, this idea leads to a crucial point. The results confirm that our model encapsulates the most important dynamical features responsible for growth towards scale independence. These constraints appear to clearly define the possible paths that a system can follow. A snapshot of its present state is then sufficient to determine where it comes from and where it is heading. This naturally leads to a second question: can we use the \emph{reconstructed} past of a system to \emph{predict} its future?

\subsection{Predicting the future}

\begin{figure*}
\begin{center}
\includegraphics[trim = 0mm 0mm 0mm 0mm, clip, width=0.82\linewidth]{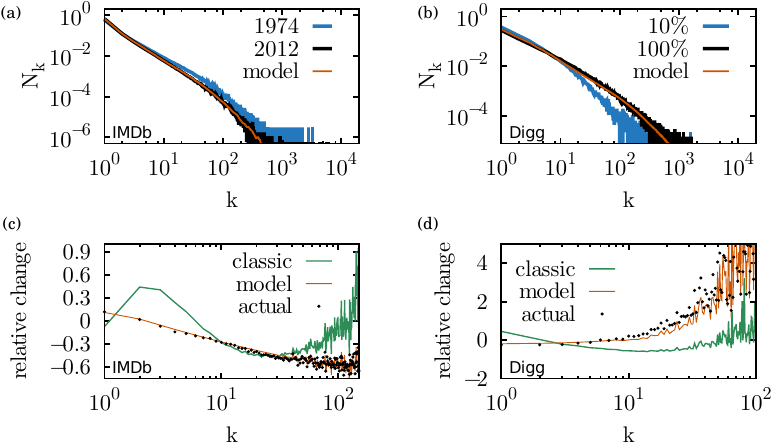}
\caption{(color online) \label{FutureResults}The model uses only the distribution at $t_i=0.3 t_f$ (IMDb) and $t_i=0.1t_f$ (Digg) of the system's history (in blue) to reconstruct the past (i.e. the birth and growth functions) and predict the future (in orange) of the database (in black). (a and b) Past, present (actual and predicted) distributions. (c and d) Relative change of each compartment $N_k$ measured as $\left[N_k(t_f)-N_k(t_i)\right]/N_k(t_i)$; where $N_k(t_f)$ is either the actual distribution or a prediction. For comparison, a prediction using the classic preferential attachment model \cite{Simon, Hebert2011_prl}, with a linear $G(k)=k$ and a time-independent $p(t) = \langle p(t) \rangle$, is shown in green.}
\end{center}
\end{figure*}

To turn our model into a predictive tool is a simple matter. We first eliminate the statistical fluctuations present in the reconstructed growth function. It is reasonable to assume that these fluctuations stem not from the form of the growth function itself but merely from the system's finite size and the stochastic nature of the dynamics. The fluctuations are eliminated by applying a linear fit to the asymptotic behavior of the reconstructed $G(k)$. A prediction can then be obtained by iterating Eq.~(\ref{model}) from a chosen present state to a desired future time.

%In order to extrapolate the results of Fig. \ref{PastResults} to a predictive model, we first eliminate any fluctuations in the growth function as these are due to finite size. Once a smoothing or fitting procedure has been applied, a prediction can be obtained by iterating the model (\ref{model}) from a chosen present state to a desired future time.

We apply this predictive model to the largest databases, i.e. actor productivity in the IMDb and user activities on Digg. The results are shown in Fig.~\ref{FutureResults}(top). By using the activity distribution on Digg after only three days (again without any temporal data, only the current activity distribution per user), we can extrapolate the distribution over the period of a month. In contrast, assuming a constant birth rate (as in classical preferential attachment \cite{Simon, Barabasi1999, Hebert2011_prl}) leads to a predicted final population of 475,000 users. Our model correctly compensates for repeated traffic and predicts a population of 115,000 users, closer to the correct value of 139,000 and missing only some sudden bursts of new user influx. This observation embodies the strength of our model and the importance of a time dependent birth rate. Similar results are obtained for actor productivity on the IMDb. Remarkably, we reproduce the state of the system at year 2012 from its state at year 1974. Given that extrapolation is a delicate procedure, it seems not unlikely that these agreements are not coincidental. As a comparison, the classical preferential attachment model shown in Fig.~\ref{FutureResults}(bottom) is incapable of discerning whether the scaling exponent of a system is increasing or decreasing with time. Since the classic model ignores the temporal dependency introduced here, our results highlight the importance of linking the temporal and organizational features of complex systems.

\begin{figure*}
\begin{center}
\includegraphics[trim = 0mm 0mm 0mm 0mm, clip, width=\linewidth]{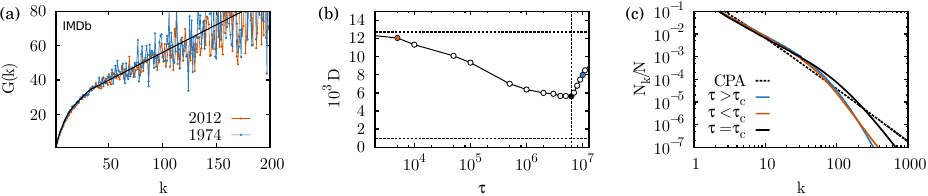}
\caption{(color online) \label{PastRobustness}(a) The growth function inferred on the full IMDb dataset (orange), as shown in Fig.~\ref{PastResults}, is compared with the function inferred with $30\%$ of IMDb's history (blue) as used in Fig.~\ref{FutureResults}. The black curve is the smooth version used to predict IMDb's future. (b) The smooth growth function of IMDb is used with different $p(t)$ to obtain distributions and measure their distance to a true power-law behavior. The lower the distance, the closer the model is to scale independence. The upper horizontal dotted line corresponds to $p(t) = \langle p(t) \rangle$ with IMDb's smooth growth function. The lower horizontal dotted line corresponds to classical preferential attachment: $p(t) = \langle p(t) \rangle$ and $G(k) = k$. With IMDb's growth function, the minimum distance (the most power-like behavior) is indicated with the vertical dotted line at $6.25\times 10^6$ ($\pm 2.5\times 10^5$) in close agreement with the MLE values of Table \ref{table1}. (c) Examples of the distributions obtained with different values of $\tau$ are compared to the classical preferential attachment (CPA) which ignores the system's intrinsic $G(k)$ by using $G(k)=k$. The color code follows the color coded dots of the middle figure.}
\end{center}
\end{figure*}

It could be argued that the growth function should more generally depend on time to include potential changes in mechanisms. However, our ability to predict the future with a time-independent growth function seems to rule out, at least in the cases studied, the necessity for a temporal dependence. In fact, Fig.~\ref{PastRobustness}(a) compares the growth function inferred from the IMDb using only records before 1974 and before 2012. While the dataset has more than tripled in size during these 40 years, the inferred growth functions do not significantly differ from one another, thereby explaining the quality of our results shown in Fig.~\ref{FutureResults}. This also implies that although the growth function has an influence on the time dependency of the dynamics (through the coupling parameter, or delay, $\tau$), it does not itself depend on time. This is particularly surprising considering that the movie industry has changed dramatically between these two snapshots. One recalls that 1975 saw the rise of the Blockbuster era following the release of Steven Spielberg's Jaws \cite{Neale03}. The following change in movie making did not affect the dynamics of the system, which suggests that the growth function may be intrinsic to the considered human activity and robust to environmental or societal changes. The growth functions of the other systems are similarly robust through time as those datasets only span between a few weeks to a few years of activity. While generalizations of our model could be considered, with growth functions varying in time or across individuals \cite{Bianconi}, the apparent time independence of the growth function is surely worthy of future investigations. Contrariwise, were the mechanism(s) of a system growth function to change over time, this would reflect immediately in our inability to predict the future and would precisely be an indication of changes in the underlying mechanism(s). Hence, even if it was to fail, this model would offer significant insights.

\subsection{Coupling of the growth function and the temporal delay}

An important insight of the previous analysis states that the delay $\tau$ embodies an inherent coupling between the growth function $G(k)$ and the birth function $p(t)$ to ensure robust scale independence.  Put differently, any  non-linearity of $G(k)$ for small $k$ should be compensated by the temporal delay $\tau$ if the system is to be roughly scale-independent even for small time $t$.

In order to examine this assertion, we make the following experiment. We use IMDb's growth function, since it is highly non-linear for small $k$, and test the plausibility of a power law fit to the model for different $p(t)$. We fix the temporal scaling $\alpha$ to IMDb's $0.55$, and we fix the value of $a$ and $b$ by setting both $p(1)$ and the average $\langle p(t) \rangle$ (for $t \in [1,5\times 10^6]$) also to that of IMDb. The only parameter allowed to vary freely is the temporal delay $\tau$. Hence, we always have the same population growing with the same growth function for the same number of time steps, and starting with the same initial birth rate but with different delays $\tau$ between the initial and the final regime of $p(t)$.

We then iterate Eq.~\ref{model} with each $p(t)$ to obtain the distribution $N_k/N$ from which we randomly generate ten populations of size $N(t)$ to emulate a real system of finite size. The generated data is then fitted to a power-law distribution with the method of Clauset, Shalizi and Newman \cite{Clauset2007}. The quality of the power-law hypothesis is finally measured with the distance between the fitted power-law distribution $N^*_k/N$ and the original distribution $N_k/N$ obtained from the model. This distance $D$ is calculated through the Jensen-Shannon divergence of the two distributions and averaged over the ten generated populations, see Appendix \S 7 for details. This approach provides an estimate of how surprising it would be for a sample obtained from our distributions to have been produced by an actual power-law distribution. %This entire procedure is schematized in Fig.~\ref{schema}.

The results highlight that, given IMDb's growth function, the particular $p(t)$ which was observed in the temporal data of IMDb and obtained from our algorithm is the most robust way for this system to grow towards scale independence. In other words, the $p(t)$ observed in the IMDb effectively compensates the non-linear deviation observed in its growth function in a way that ensures a fast convergence to scale independence. Figure \ref{PastRobustness}(c) illustrates this point by comparing three distributions obtained with different $p(t)$ with the classical preferential attachment ($[p(t) = <p(t)>, G(k) = k]$). The distribution obtained with the optimal solution ($\tau = \tau_c$) is clearly ahead of the other, and not so far from the CPA, on the path to scale independence.

To intuitively interpret these results, one can think of the need to populate both the linear and non-linear regime (if any) of the growth function $G(k)$ to obtain true scale independence. This can be done in one of three ways: either by (i) building up population in the non-linear regime then allowing it to grow into the linear regime ($p(t)$ decreasing); (ii) bringing early individuals to the linear growth regime then build up population in the non-linear regime ($p(t)$ increasing); or (iii) continuously balancing between birth and growth events ($p(t)$ constant). For instance, the form of IMDb's growth function explains its increasing birth function: one should quickly create a population with large $k_i(t)$ (in the linear regime of $G(k)$) rather than build up population density in the non-linear regime. Otherwise, this population would take a long time to move towards the linear regime because of the diminishing returns of growth ($d^2G(k)/dk^2 < 0$). Yet, enriching a population fraction with unnecessary large shares $\{ k_i(t) \}$ before building up the population with small shares would obviously also slow down the emergence of scale independence. This trade-off explains the existence of an optimal delay as observed in Fig.~\ref{PastRobustness}(b).

In a nutshell, this simple experiment adds further strength to the validity of our theoretical framework, and reasserts one of its important conclusions: arbitrary growth rules do not all lead to scale independence, and  certainly not all at the same speed. Finally, while we have confirmed our theoretical insights and our ability to use them in practical applications, the mechanisms by which $p(t)$ might self-organize in these systems to assure scale independence remain unknown.

\section{Conclusion\label{sec:conc}}

In this paper, instead of directly studying the classical preferential attachment model, we have derived a more general form from the simple assumption that a power-law distribution is a good \textit{approximation} of a distribution of interest. Our general model differs from the classic idealized version in two ways: the growth (or attachment) function is given some flexibility in its initial behavior, only required to be asymptotically linear; and the birth function is \textit{time dependent} through a delayed temporal scaling. While only the constraint on the growth function is necessary to converge towards scale-free organization, the time dependent birth function can compensate non-linearity in growth and hasten the system's convergence through a delayed temporal scaling. This delay acts as a coupling between two levels of dynamics: the growth of the population and the growth of a given individual's activity.

This general model is both flexible and constrained enough to be useful. In fact, we have shown that a three dimensional parameter space (temporal scale exponent, delay and asymptotic birth rate) is sufficient to capture the time dependency of a present distribution.

It is important to keep in mind that our analysis is in no way restricted by the nature of the systems under study. Considering that scale independent systems are ubiquitous in science and everyday life, but that temporal data on their growth is seldom available, our framework provides a new investigation line to reconstruct their past and to forecast their future.

\begin{acknowledgments}
The authors would like to acknowledge Calcul Qu\'{e}bec for computing facilities, as well as the financial support of the Canadian Institutes of Health Research, the Natural Sciences and Engineering Research Council of Canada, the Fonds de recherche du Qu\'ebec--Nature et technologies and the James S. McDonnell Foundation. 
\end{acknowledgments}

\appendix*

\section{Data and methods\label{methodssec}}

\subsection{Derivation of Eq.~\eqref{kmax} : the slope $\kappa$ and the delay $\tau$}
The derivation is based on the following arguments.
Let $G(k) \propto k$ $\forall\ k \geq k^*$.
Without loss of generality, the slope of the linear behavior is taken to be equal to one. 
Let us then write 
\begin{align}
    S_G= &\; \sum_i G(k_i(t)) = \sum_{k=1}^{k_m(t)} G(k) N_k(t) \nonumber \\ = &\: \sum_{k=1}^{k^*-1} G(k) N_k(t) +  \sum_{k=k^*}^{k_m(t)} k N_k(t).
\end{align}
With our definition of time $t$
\begin{equation}
    \sum_{k=1}^{k_m(t)} k N_k(t)   = \left[ \sum_{k=1}^{k^*-1} + \sum_{k=k^*}^{k_m(t)} \right] k N_k(t) = t \;,
\end{equation}
we can combine the summations to obtain
\begin{align}
    S_G = &\; t + \left\{\sum_{k=1}^{k^*-1} \left(G(k) - k\right)N_k(t)\right\} \nonumber \\ \equiv &\: t + \left\{\sum_{k=1}^{k^*-1} \Delta G(k) N_k(t) \right\} \;.
\end{align}
For large enough times, we may assume that $\{ N_k(t)\}$ has reached its stationary distribution,  $N_k(t) = A^* k^{-\gamma} N(t)$ (see Eq.~\eqref{k_from_N} of the main text).
The previous equation then simplifies to
\begin{equation}
    S_G = t +  N(t)  \Delta G^* ,    
\end{equation}
where $\Delta G^*$ is a constant quantifying the (weighted) deviation at small $k < k^*$ between the actual $G(k)$ and its linear asymptotic behavior.
The next step involves the separation of $p(t)$ into a time-dependent  and an asymptotic part, $f(t)$ and b respectively,
\begin{equation}
    p(t) = f(t) + b .
\end{equation}
Since $p(t) \in [0,1]$, $f(t)$ is bounded to the interval $[-b, 1-b]$.
Furthermore, 
because $p(t)$ is the time derivative of the total population, $p(t)= \dot{N}(t)$, integration leads to 
\begin{equation}
N(t) = N(1) +\int_1^t p(t') dt' = 1 + \left[ F(t) + b(t-1) \right]
\end{equation}
   
 and $S_G$ becomes
\begin{equation}
  S_G = \left[1 +   b \Delta G^* \right] t    +  \left[1 - b +F(t) \right] \Delta G^*\equiv \kappa \left[ t + \tau(t) \right]    
  \label{eq:SofG}
\end{equation}
 with 
 
\begin{equation}
   \kappa = (1 +  b \Delta G^*) \label{eqn:kappa}
\end{equation}
  and 
\begin{equation}
\tau(t) =    \left[1 - b +F(t) \right]  \Delta G^* / \left[1 +   b \Delta G^* \right]  \ . \label{eqn:tauoft}
\end{equation} 
   
The constant $\kappa$ will only be equal to one if $b=0$ and / or $\Delta G^*=0$.
What is left to investigate is the time dependence of $\tau(t)$.\\

\noindent{\sc Case 1:} $p(t) = b $ and $f(t)=0$.
This is the simplest case where $\tau(t)= \tau_1 = \left[1 - b  \right]  \Delta G^* / \left[1 +   b \Delta G^* \right]$.\\

\noindent{\sc Case 2:} $p(t) = f(t) + b$ with $b \not= 0$.
Since the asymptotic growth of $N(t)$ will be dominated by the term $b t$, the integral $F(t)$ will converge to a constant, say $F^*$, leading to a constant delay $\tau(t) \simeq \tau_2 =  \left[1 - b +F^* \right]  \Delta G^* / \left[1 + b \Delta G^* \right]$.\\
                        
\noindent{\sc Case 3:} $p(t)= f(t)$ and $b=0$.
There is a remaining time dependence from  $F(t)$, but this is correct since it is responsible for the growth of $N(t)$ at large times.
However, we have established in Eq. (12) , a relationship between $p(t)$ and $k_{\rm max}(t)$, and since $k_{\rm max}(t)$ grows as $t^\delta$ ($0<\delta<1$), whatever the precise value of this exponent, the result is a {\em sub-linear} growth for $F(t)$,
i.e. $\tau(t) \simeq \tau_3 =  \Delta G^* + {\cal O}(t^\eta)$ ($\eta < 1$).
For large enough $t$, the extra time dependence can be safely discarded in front of the linear term of Eq.~\eqref{eq:SofG}.\\

To summarize, in all cases,
\begin{equation}
    S_G= \kappa (t + \tau) .
\end{equation}
Eq.~\eqref{kmax} arises then from Eq.~\eqref{k} of the main text as we follow the evolution of the leader $k_{\rm max}(t)$ beyond a certain $t \geq t^*$
\begin{align}
            k_{\rm max}(t+1) = &\:  k_{\rm max}(t)  + \left[ 1 -p(t) \right] \frac{G\left( k_{\rm max}(t) \right)}{S_G} \nonumber \\ = &\: \left[ 1 + \frac{1 -p(t)}{\kappa (t +\tau)} \right] k_{\max}(t)\; .
\end{align}
The transition to continuous time leads to the differential equation, Eq. (\ref{eq:diff_kmax}).

\subsection{Description of databases}

\textbf{Prose samples.} Text files for the works of William Shakespeare, Miguel de Cervantes Saavedra and Johann Wolfgang von Goethe were downloaded from the Project Gutenberg at \texttt{www.gutenberg.org/}. Punctuation marks and Project Gutenberg disclaimers were removed from the files manually. %The files were then broken into samples of equal length and analyzed separately.

While not a human system, but certainly a man-made one, these prose samples were used to get better statistics on the birth function. While human systems are unique and time dependent, written texts feature a translational invariance \cite{Bernhardsson2010}. This property allows us to gain better statistics of their growth by considering multiple samples of equal length as different realizations of the same process.

Time $t$ and resource $K(t)$ correspond to the total number of written words. Individuals correspond to unique words and their share $k_i(t)$ to their number of occurrences.

\textbf{Scientific authorships on the arXiv.} This database consists of a chronological list of all author names appearing on papers of the arXiv preprint archive (in order of publication date). It was compiled using the arXiv API to gain a full list of scientific publications available from \texttt{http://arxiv.org/} as of April 2012.

Time $t$ and resource $K(t)$ correspond to the total number of paper authorships. Individuals correspond to authors and their share $k_i(t)$ to their number of publications.

\textbf{Digg user activities} Digg (\texttt{http://digg.com/}) is a social news website where registered users can vote on news or other types of articles that they deem interesting. This database is a list of all user votes on top stories (frontpage) over a period of one month in 2009 \cite{Lerman10}.

Time $t$ and resource $K(t)$ correspond to the total number of votes. Individuals correspond to registered users and their share $k_i(t)$ is their respective number of votes.

\textbf{IMDb castings} The Internet Movie Database (\texttt{http://www.imdb.com/}) consists of an impressive amount of cross referenced lists (released films, cast and crew, etc.). These databases can be accessed or downloaded in various ways: see \texttt{http://www.imdb.com/interfaces} for details. From the list of actors featured on IMDb, which records all movies in which they have appeared, and the list of movie release dates, we built the chronological sequence of `castings'.

Time $t$ and resource $K(t)$ correspond to the total number of castings (a given actor playing in a given film). Individuals correspond to unique actors and their share $k_i(t)$ is the total number of films in which they have appeared.

\textbf{Sexual activities in a Brazilian community} This database was built from a public online forum for male clients who evaluate relations with female prostitutes \cite{Holme2011}. After preliminary results using the client and prostitute databases separately, we concluded that it was not necessary to distinguish between the two. The simplified database is thus a list of unique identification numbers (IDs) corresponding to either a client or a prostitute, in chronological order of sexual relations (at time of online posting).

Time $t$ and resource $K(t)$ correspond to the total number of such IDs (twice the total number of relations). Individuals correspond to unique IDs (either client or prostitute) and their share $k_i(t)$ is their respective number of relations.

\begin{table}[h!]
\caption{Summary of database sizes and quantities.}
%\vspace{-0.5cm}
\begin{center}
\begin{tabular}{ @{}l c c c c@{} } \hline\hline 
Quantities & arXiv & Digg & IMDb & Sexual\\
\hline 
\hline
Individuals & authors & users & actors & clients/prostitutes\\
$N(t_f)$ & 386,267 & 139,409 & 1,707,565 & 16,730\\
Resource & papers & votes & castings & sexual activities\\
$K(t_f)=t_f$ & 1,206,570 & 3,018,197 & 6,288,201 & 101,264\\
\hline
\hline
\end{tabular}
\end{center}
\label{table2}
\end{table}

\subsection{Measuring the birth function}

\textbf{Prose samples} The translational (or temporal) invariance of written text implies that we can consider different samples of equal length from the same author as different realizations of the same experiment. The files were thus broken into samples of equal length and analysed separately. Each experiment can be reduced to a binary sequence of ones (when the word is a new word; i.e. a birth event) and zeros (when the word is an old one; a growth event). The birth function $p(t)$ of a given author can then be obtained by simply averaging all binary sequences.

\textbf{Other systems} In the other systems, since preliminary tests excluded the possibility of temporal invariance, a different procedure was used. The simplest one is to merely apply a running average on the binary sequence of birth and growth events. We used temporal windows of $\Delta t$ equal to $1\%$ of the total system size (final time $t_f$) for the two largest databases (Digg and IMDb) and between $0.5\%$ and $1\%$ of system size for the others. This method was shown to preserve the delayed temporal scaling on a random binary sequence whose elements were drawn from a known probability distribution following $p(t)$.

\subsection{Measuring the growth function}

We now describe the procedure used to obtain the growth function $G(k)$ of a system from its temporal data, $t \in [0, t_f]$. We use the following notation: we keep in memory every encountered individual $i$, its number of appearances (or current share) $k_i(t)$, $N_k(t)$ as the number of individuals with share $k_i(t) = k$ and the total population $N(t)$ after time $t$. Starting from $t=1$, we proceed as follows.
\algblock{If}{EndIf}
\algcblock[If]{If}{ElsIf}{EndIf}
\algcblock{If}{Else}{EndIf}\algcblockdefx[Strange]{}{Eeee}{Oooo}
[1]{\textbf{Input} #1}
[1]{\textbf{Output} #1}
\begin{algorithmic}[1]
\Eeee{individual $i$ involved in event $t \in [0, t_f]$}
\Oooo{measured growth function $G(k)$}
\FORALL{$t \in [0, t_f]$}
\IF{the individual $i$ involved in the $t$-th event is new}
	\STATE{add it to memory and update:}
	 \begin{eqnarray}
	N(t) & = & N(t-1) + 1 \nonumber \\
	k_{N(t)}(t) & = & 1 \nonumber \\
	N_1(t) & = & N_1(t-1) + 1 \nonumber
	\end{eqnarray}
	\ELSE{ increment a function of chances:}
	\begin{equation}
		C(k,t) = C(k,t-1) + N_k(t-1)/N(t-1) \quad \forall \; k \; \nonumber
	\end{equation}
	\STATE{increment a function of successes:}
	\begin{eqnarray}
		S(k_i(t-1),t) & = & S(k_i(t-1),t-1) + 1 \nonumber \\
		S(k,t) & = & S(k,t-1) \qquad \forall \; k \neq k_i(t-1) \nonumber
	\end{eqnarray}
	\STATE{update the following variables:}
	\begin{eqnarray}
	k_i(t) & = & k_i(t-1) + 1 \nonumber \\
	N_{k_i(t-1)}(t) & = & N_{k_i(t-1)}(t-1) - 1 \nonumber \\
	N_{k_i(t)}(t) & = & N_{k_i(t)}(t-1) + 1 \nonumber
	\end{eqnarray}
	\ENDIF
\ENDFOR
\STATE{the growth function is:}
\begin{equation}
G(k) = S(k,t_f)/C(k,t_f) \quad \forall \; k \; \; .\nonumber
\end{equation}
\end{algorithmic}
The obtained $G(k)$ corresponds to the ratio of actual successes to chances under a uniform growth.

\subsection{Reconstructing the empirical growth function}

Once the best possible $p(t)$ has been found, we adjust the growth function $G(k)$ by iterating the following algorithm:

\algblock{If}{EndIf}
\algcblock[If]{If}{ElsIf}{EndIf}
\algcblock{If}{Else}{EndIf}
\algcblockdefx[Strange]{}{Eeee}{Oooo}
[1]{\textbf{Input} #1}
[1]{\textbf{Output} #1}
\begin{algorithmic}[1]
\Eeee{target $\tilde{N}_k$ and first approximation $G(k)=k$}
\Oooo{adjusted growth function $G(k)$}
\STATE{initial condition $N_k(1) = \delta _{k1}$}
\FORALL{$t \in [0, t_f]$}
	\begin{align}
N_k(t+1) = &N_k(t)+ p(t)\delta _{k1} \nonumber \\ &+ \frac{1- p(t)}{\sum G(k)N_k(t)}\left[G\left(k-1\right)N_{k-1}(t) -  G(k)N_k(t)\right] \; . \nonumber
	\end{align}
\ENDFOR
\FORALL{$k \in [0, k_{\textrm{max}}(t_f)]$}
\begin{equation}
\overline{G}(k) = G(k)\frac{N_k(t_f)/\sum _{i=k}^{\infty} N_i(t_f)}{\tilde{N}_k(t_f)/\sum _{i=k}^{\infty} \tilde{N}_i(t_f)} \nonumber
\end{equation}
\ENDFOR
\STATE{set $G(k) = \overline{G}(k)$.}
\end{algorithmic}
At step 6, the adjustment factor is simply the ratio of ``the quantity of individuals (water) that made it to share (bucket) $k$ but did not go to $k+1$'', as calculated in the model $N_k(t_f)$ versus the target distribution $\tilde{N}_k(t_f)$. This algorithm is usually iterated 4 or 5 times to obtain a converged growth function.

\subsection{Maximum-likelihood estimation}
We search for a $p(t)$ that maximizes the binary logarithm of the likelihood $\mathcal{L}$ of a given binary sequence $\{y_i\}$ of birth ($y_i = 1$) and growth events ($y_i = 0$):
\begin{equation}
\log_2 \mathcal{L}\left(\tau, \alpha, b \;\vert\; \lbrace y \rbrace\right) = \sum _{i=1}^{t_f} y_i \log_2 p(i) + \left(1-y_i\right)\log_2 \left(1-p(i)\right) \; . \nonumber
\end{equation}

\subsection{Jensen-Shannon divergence}

Given two distributions, $\bm{M}$ and $\bm{F}$, with probabilities $\{M_i\}$ and $\{F_i\}$ respectively, the quantity
\begin{equation}
D_{\textrm{KL}}\left(\bm{M}\Vert\bm{F}\right) = \sum _i M_i\log_2 \left(\frac{M_i}{F_i}\right)
\end{equation}
is called the \textit{Kullback-Leibler distance} \cite{numerical_recipes} between $\bm{M}$ and $\bm{F}$, or the relative entropy between the two distributions. A close relative of this quantity, also referred to as the \textit{Jensen-Shannon divergence}, is a symmetric form given by
\begin{equation}
D_{\textrm{SKL}} = \frac{1}{2}D_{\textrm{KL}}\left(\bm{M}\Vert\bm{A}\right) + \frac{1}{2}D_{\textrm{KL}}\left(\bm{F}\Vert\bm{A}\right)
\end{equation}
where the distribution $\bm{A}$ with probabilities $A_i = \left(M_i+F_i\right)/2$ is used to approximate $\bm{M}$ or $\bm{F}$ respectively.

In our study, we want to quantify the similarity between the distribution, $\bm{M}$, generated by our mean-field model and the distribution $\bm{F}$ obtained from a corresponding power-law fit. In practice, the procedure goes as follows: with the distribution $\bm{M} = \{N_k/N\}$, we generate a number of population samples $\{m^{(j)}\}$ of size $N(t_f)$ and fit each of them to a power-law $f^ {(j)}$ using the standard method of Clauset \textit{et al}. \cite{Clauset2007}. Each $f^{(j)}$ is characterized by an exponent $\gamma ^{(j)}$ and a minimal value $k_{\textrm{min}}^{(j)}$ (here always equal to 2) marking the beginning of the power-law tail. These power-law populations are then used to construct the related distributions $\left[\bm{F}^{(j)} = \{N_k^{(j)}/N\}\right]$ which are finally compared to the tail of the original distribution $\bm{M}$ over the range $k_{\textrm{min}}^{(j)} \leq k \leq 5000$ [$\sim$ IMDb's $k_{\textrm{max}}(t)$]. The comparison is quantified through the symmetrical Kullback-Leibler distance averaged over the different samples
\begin{equation}
D\left(\bm{M},\bm{F}\right) = \langle D_{\textrm{SKL}}\left(\bm{M},\bm{F}^{(j)}\right)\rangle _j .
\end{equation}

\color{black}

\end{document}